# Employing Spectral Domain Features for Efficient Collaborative Filtering

Doaa M. Shawky[1]

**ABSTRACT**

Collaborative filtering (CF) is a powerful recommender system that generates a list of recommended items for an active user based on the ratings of similar users. This paper presents a novel approach to CF by first finding the set of users similar to the active user by adopting self-organizing maps (SOM), followed by k-means clustering. Then, the ratings for each item in the cluster closest to the active user are mapped to the frequency domain using the Discrete Fourier Transform (DFT). The power spectra of the mapped ratings are generated, and a new similarity measure based on the coherence of these power spectra is calculated. The proposed similarity measure is more time efficient than current state-of-the-art measures. Moreover, it can capture the global similarity between users' profiles. Experimental results show that the proposed approach overcomes the major problems in existing CF algorithms as follows: First, it mitigates the scalability problem by creating clusters of similar users and applying the time-efficient similarity measure. Second, its frequency-based similarity measure is less sensitive to sparsity problems because the DFT performs efficiently even with sparse data. Third, it outperforms standard similarity measures in terms of accuracy.

## 1. INTRODUCTION

Recommender systems (RS) have been increasingly adopted in online applications to provide users with enhanced services and a better web surfing experience. In RS, an item is usually recommended to users based on their ratings of similar items. Collaborative filtering (CF) [1] is a commonly used technique in RS that has been employed effectively in e-commerce [2]. Collaborative filtering generates recommended items based on other users' opinions. This technique generates recommended items for an active user by analyzing other users' ratings of a set of items. This set is called the rating matrix. Each element $r_{ij}$ in the rating matrix represents a rating provided by user $i$ regarding item $j$.

CF algorithms can be divided into two main categories depending on how the rating matrix is processed. These categories include memory-based and model-based algorithms. In the former approach, a similarity measure is applied to the entire rating matrix to generate a set of users (or items) similar to the active user. In the latter approach, a model is built using the rating matrix to simulate the user's behavior; hence, the active user's preferences can be predicted using that model. The model-based approaches primarily adopt machine learning and use linear algebra theories to generate a model and select its parameters.

[1] Engineering Mathematics Department, Faculty of Engineering, Cairo University, 12613, Cairo, Egypt (email: doaashawky@staff.cu.edu.eg)

In comparison with memory-based approaches, model-based approaches are less sensitive to the common problems of CF, namely, the small number of ratings that are often available for items (the sparsity problem) and the lack of reliable recommendations for newly added users and/or items in the rating matrix (the cold-start problem) [2]. In contrast, memory-based approaches are much simpler to implement and—unlike the model-based approaches—do not suffer from parameter selection and optimization problems [2-4]. To overcome the problems of both model-based and memory-based approaches, a combination of the two is sometimes used. For example, instead of finding similar users using the full rating matrix, a group of similar users is first found and then items are recommended to the active user using only those users who are in the group closest to the active user. However, there is a tradeoff between the scalability of these approaches and the prediction accuracy [3].

This paper presents an approach for an effective CF technique using a combined model-based and memory-based algorithm. The approach adopts two powerful clustering techniques to generate clusters of similar users: self-organizing maps (SOM) and k-means clustering. Because SOM can find hidden similarities in data, it is logical to expect that applying SOM followed by k-means clustering can enhance the clustering efficiency. Furthermore, to find the most similar items to an active user's preferences, first, the closest cluster to the active user is found. Then, the ratings of each item provided by all users in that cluster are mapped to the frequency domain using the Discrete Fourier Transform (DFT). Then, using the power spectra of the items' representations, a novel similarity measure based on coherence is defined. In addition to the novelty of the proposed approach, it yields high prediction accuracy when compared to those that apply standard similarity measures.

The rest of the paper is organized as follows. Section 2 introduces a survey of the related research. The background of some related concepts is presented in Section 3. In Section 4, the proposed approach is presented in detail. Section 5 presents the performed experimental study and discusses the results. Finally, Section 6 presents conclusions and highlights some directions for future work.

## 2. RELATED RESEARCH

The literature contains numerous studies regarding RS, particularly in regard to CF. Thus, it is not possible to provide a comprehensive list of those approaches. However, in this section, a review of the most recent approaches will be presented.

In [5], for instance, an approach for clustering users based on their rating patterns, followed by a combined rule-based and case-based reasoning, is proposed. The MovieLens 100K [6] dataset is used to show that the proposed approach outperforms common CF models using standard similarity measures such as the Pearson similarity measure [7]. Also, in [8] presents an approach that addresses the sparsity problem using Bhattacharyya similarity. The approach utilizes the numerical values of all ratings made by a pair of users, not just the ratings submitted for common items. Thus, it combines local and global similarity to obtain the final similarity value between a pair of users. Experiments using MovieLens, Netflix, and Yahoo datasets showed that this approach outperformed other approaches in terms of the evaluation metrics used. Similarly, in [9], Haifeng et al. proposed a new heuristic similarity measure based on three factors—proximity, impact and popularity—to overcome some of the typical problems with commonly used measures. The proximity factor represents how closely the ratings agree, penalizing ratings in disagreement. The impact factor represents how strongly an item is preferred or disliked. Finally, the popularity factor denotes how common two users' ratings are. The experiments in this study used two versions of the MovieLens datasets (ML-100K and ML-1M) and the Epinions dataset. The results showed the effectiveness of the proposed similarity measure in comparison to other well-known methods. Wasid and Kant [10] applied fuzzy sets to represent the imprecise features of users. Moreover, they applied Particle Swarm Optimization [11] to learn user-specific weights for various features. Experiments performed using the MovieLens dataset showed the effectiveness of the proposed approach in comparison with Pearson-based CF systems and fuzzy genetic systems. Also in [12], an incremental CF approach based on Mahalanobis distance [13] was proposed. This approach builds clusters of similar users using an incremental algorithm based on the Mahalanobis radial basis function. The prediction phase employs fuzzy sets to determine the membership degree of each user to the formed clusters. The MovieLens dataset was used to show the effectiveness of the proposed approach. In [14], the authors presented an approach that deals with the sparsity and computational issues of CF systems by constructing nonredundant subspaces that represent each user. Then, a tree is built to represent these users' commonality with the active user, followed by a new recommendation method to predict unknown ratings. Experiments performed using the MovieLens and Jester datasets showed that the proposed approach was effective in most cases. Moreover, in [15], an interest intensity model that decayed over time was built to represent how users' preferences are correlated and change with time. A scoring matrix is then constructed that can alleviate problems with the sparse user-item scoring matrix. Also, in [16], a multi-criteria approach for representing imprecise aspects of user behavior with respect to items is introduced. The approach applies the Adaptive Neuro-Fuzzy Inference System combined with subtractive clustering and Higher Order Singular Value Decomposition to deal with imprecise and high-dimensional data. These experiments used the Yahoo!Movies dataset and showed improved prediction accuracy compared to similarity-based approaches. A reversed CF was also presented in [17]. This approach utilized a k-nearest neighbor (k-NN) graph, which, instead of finding k similar neighbors of unrated items, finds the k-nearest neighbors of rated items. Experimental results using the MovieLens dataset show that this approach outperforms traditional user-based/item-based CF algorithms in terms of preprocessing and query processing time without sacrificing the level of accuracy. Also, in [18], a mulit-level recommendation method that modifies similarities between users using Pearson Correlation Similarity based on some constraints is proposed. The effectiveness of the presented approach is tested on three datasets.

## 3. DEFINITIONS OF RELATED TERMS
### 3.1 SPECTRAL DOMAIN FEATURES

For a full understanding of the analogy between digital signals and CF, as proposed in the presented approach, this section provides a brief description of some digital signal processing (DSP) techniques.

Basically, DSP represents a signal using a sequence of numbers [18]. The numbers can then be manipulated or altered by a computing process to change or extract information from the original signal. A signal is denoted by $x[n]$, which is a sequence of numbers in which $n$ represents the sample number and $x[n]$ is the sample value. We can use a method such as the Fourier Transform (FT) to find the frequency components in a signal. Each component is called a harmonic component. Thus, every signal has a frequency spectrum [20].

Many sequences can be represented by the DFT, as given by (1) [21]:

$$X(e^{i\omega}) = \sum_{n=-\infty}^{\infty} x[n]e^{-i\omega n} \quad (1)$$

and

$$x[n] = \frac{1}{2\pi} \int_{-\pi}^{\pi} X(e^{i\omega})e^{i\omega n}\, d\omega, \quad (2)$$

where (1) denotes the DFT of the sequence $x[n]$, and (2) denotes the inverse DFT. For simplicity, the DFT of $x[n]$ will be denoted by $X(\omega)$. The DFT determines how much of each frequency component over the interval $-\pi \leq \omega \leq \pi$ is required to synthesize $x[n]$. Moreover, the DFT is usually referred to as the spectrum of the sequence $x[n]$. Practically, the Fast Fourier Transform (FFT) is usually used, which is fundamentally an efficient implementation of the DFT with less complexity. Note that for a dataset with n points, the time complexity for calculating FFT is O(nlogn) [22]. Also, only a few coefficients of the FFT are required to reconstruct the sequence x[n], which makes using FFT to analyze a sequence a time-efficient process.

Coherence between two signals, $x[n]$ and $y[n]$, is "a measure of the degree of relationship, as a function of frequency, between the two series" [23]. It is usually denoted by $C_{xy}$ and is given by (3).

$$C_{xy} = \frac{\Phi_{xy}(\omega)}{\sqrt{\Phi_{yy}(\omega)\Phi_{xx}(\omega)}} \quad (3)$$

where $\Phi_{xy}(\omega) = Y(\omega)X^*(\omega)$ is the cross-power density between X($\omega$) and Y($\omega$) and $X^*(\omega)$ is the conjugate of X($\omega$). Moreover, $\Phi_{xx}(\omega) = X(\omega)X^*(\omega)$ which is the auto-power spectrum (or simply, the power spectrum) of X($\omega$). The coherence $C_{xy}$ satisfies the property $|C_{xy}| \leq 1$. Moreover, if the coherence vanishes in some frequency band, the lack of coherence means that there is no relationship between $x[n]$ and $y[n]$ in that band. Thus, spectral coherence identifies frequency domain correlation between signals, where small coherence values indicate that the corresponding frequency components are uncorrelated. On the other hand, values tending towards 1 indicate that the corresponding frequency components are correlated.

In the context of CF, a rating matrix is used. The values in each cell of this matrix are represented by $r_{ij}$, where $i = 1,2,..,m$ ($m$ is the number of users in the system) and $j = 1,2,..,n$ ($n$ is the number of items in the system). To be able to treat this matrix as a digital signal, the sequence of values in each row $i$, which represents the ratings user $i$ has provided for all items, can be used. On the other hand, the sequence of values in each column can also be used. In this case, the used sequence represents the ratings provided for item $j$ by all users in the system. Thus, coherence between sequences represented by two different columns can be used to represent how similar these two columns are.

## 3.2 SELF-ORGANIZING MAPS (SOM)

With analogy to the human brain, where similar information tends to reside in close proximity, the idea behind the SOM is that neurons with similar weights should be grouped together because they represent similar outputs [24]. As shown in Fig. 1, SOM have a feed-forward structure with a single computational layer arranged in rows and columns. Each neuron is fully connected to all the source nodes in the input layer.

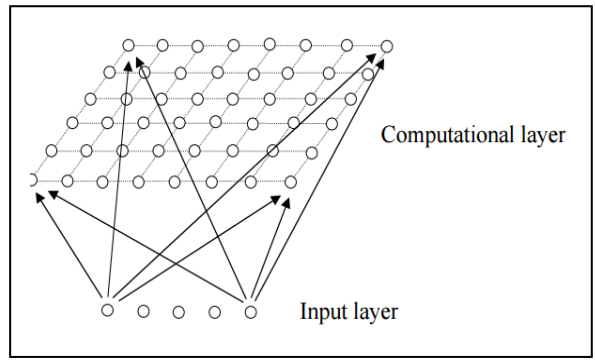

Fig. 1. Architecture of SOM with 2-dimensional lattice.

The competitive learning algorithm used in SOM can be intuitively divided into the following four stages. First, the nodes' weights are randomly initialized. Second, an input is randomly chosen from the training data. Third, a winning neuron is selected to be the one with the closest weights to the input vector. Fourth, the neighborhood of the winning neuron is randomly initialized and then is iteratively decreased to include only the neurons that are close enough to the winning neuron. Each weight vector then moves to the average position of all input vectors for which it is a winner or for which it is in the neighborhood of a winner. The distance that defines the size of the neighborhood is altered during training through two phases: the ordering and tuning phases. In the ordering phase, the neighborhood size is gradually reduced until it reaches a certain value and the neurons order themselves in the input space with the same topology in which they are ordered physically. In the second phase, the neighborhood size is further decreased below that defined value and the weights are expected to spread out relatively evenly over the input space while retaining their topological order found during the ordering phase. Thus, the neurons' weight vectors initially take large steps all together toward the area of the input space where the input vectors are occurring. The training continues to give the neurons time to spread out evenly across the input vectors. If the input vectors occur with varying frequency throughout the input space, the computational layer (feature maps) tends to allocate neurons to an area in proportion to the frequency of the input vectors found there. Thus, feature maps, while learning to categorize their input, also learn both the topology and distribution of their input. Note that SOM differs from conventional competitive learning in terms of which neurons' weights are updated. Instead of updating only the winner, feature maps update the weights of the winner and its neighbors. The result is that neighboring neurons tend to have similar weight vectors and, thus, are responsive to similar input vectors [25].

More formally, let the input vector be denoted by $x_i$, i = 1, 2,…, D, where D is the dimension of the input space. If we use a topology with N neurons, we can then define the weights that connect neurons to the input vectors as $w_{ij}$, where j = 1, 2, …, N. We can find the neuron closest to the input by calculating the Euclidean distance, for example, between the input vector $x$ and the weight vector $w_j$ for each neuron j, as follows: [26]

$$d_j(x) = \sqrt{\sum_{i=1}^{D}(x_i - w_{ij})^2}. \quad (4)$$

Finding the winning neuron will result in a set of excited neurons that belong to the neighborhood of the winning one. If $d_{ij}$ is the lateral distance between neuron $i$ and neuron $j$, the neighborhood $T_{j,I(x)}$ between a winning neuron $i$ and a neighboring neuron $j$ could be calculated as shown in (5)

$$T_{j,I(x)} = e^{\left(-\frac{d_{j,I(x)}^2}{2\sigma^2}\right)}, \quad (5)$$

where $I(x)$ is the index of the winning neuron. Note that this function is maximum at the winning neuron. Moreover, it decreases to zero as the distance increases to infinity. To enable the neighborhood function to decrease with time, σ is usually an exponential function that decays with time, as presented in (6)

$$\sigma(t) = \sigma_0 e^{\left(-\frac{t}{\tau}\right)}, \quad (6)$$

where τ is the learning rate.

Moreover, during the learning phase, the weights are updated using (7)

$$\Delta w_{ij} = \sigma(t) T_{j,I(x)} (x_i - w_{ij}) \quad (7)$$

### 3.3 K-MEANS CLUSTERING

One common, easy-to-implement clustering algorithm is k-means [27]. Informally, k-means clustering starts with k points as arbitrary initial centroids, where k is a user-specified parameter. Each data point is then assigned to the cluster with the closest centroid. Then, the centroid of each cluster is updated by taking the average of the data points that belong to each cluster. The process is repeated until all data points in each cluster are within a specified distance to the centroid. One of the drawbacks of k-means is that the clustering result depends on the value of k [27].

### 4. PROPOSED APPROACH

## 4. 1 PROPOSED SIMILARITY MEASURE

Similarity between users has been commonly calculated using some measures such as the Pearson correlation coefficient, Jaccard similarity measure [28], and several other measures proposed in the literature. A list of the state-of-the-art measures can be found in (9). Note that the time complexity for calculating these measures is $O(n^2)$, where n is the number of users. For example, the Pearson correlation coefficient (PCC) can be calculated using (8)

$$PCC(u,v) = \frac{\sum_{P \in I}(r_{up} - \overline{r_u})(r_{vp} - \overline{r_v})}{\sqrt{\sum_{P \in I}(r_{up} - \overline{r_u})^2} \sqrt{\sum_{P \in I}(r_{vp} - \overline{r_v})^2}}, \quad (8)$$

where $I$ represents the set of items that are common to users $u$ and $v$, and $\overline{r_u}$ and $\overline{r_v}$ are the average ratings provided by user $u$ and $v$, respectively. Additionally, the Jaccard similarity measure is defined by (9)

$$Jaccard = \frac{|I_u| \cap |I_v|}{|I_u| \cup |I_v|}, \quad (9)$$

where $I_u$ and $I_v$ are the sets of items recommended by users $u$ and $v$, respectively.

Other similarity measures that will be used in this work include the Mean Squared Difference (MSD) [29] and Jaccard MSD [30].

To illustrate how the new similarity measure is calculated, consider the three sequences $x[n]$, $y[n]$, and $z[n]$, each of which has 20 real values. As shown in Fig. 2, the x sequence is more similar to the y sequence than it is to the z sequence. The spectra of the three sequences are also shown in the figure. Note that the spectrum of y is more similar to the spectrum of x than it is to the spectrum of z. Moreover, Fig. 3 shows the coherence between x, y and x, z at different frequency values. As shown, the number of frequencies with high coherence between x and y is larger than that between x and z. Thus, based on coherence, we can define the new similarity measure $CohrSim(x, y)$ as given by (10)

$$CohrSim(x,y) = \frac{\sum_f C_{xy}}{length(C_{xy})}, \quad (10)$$

where $f$ is the frequency band over which $C_{xy}$ is calculated and $length(C_{xy}) = n$, if $C_{xy}$ is a sequence of $n$ values.

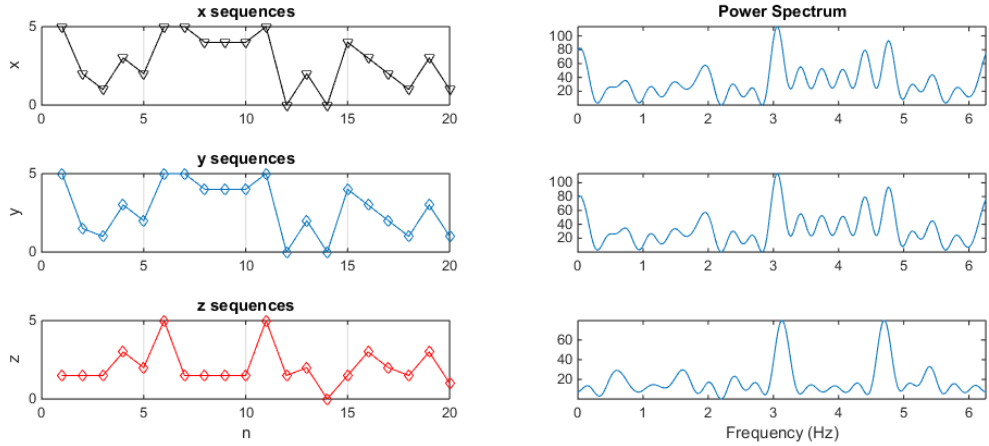

Fig. 2. The x, y, and z sequences (left) and the power spectra of x, y, and z (right).

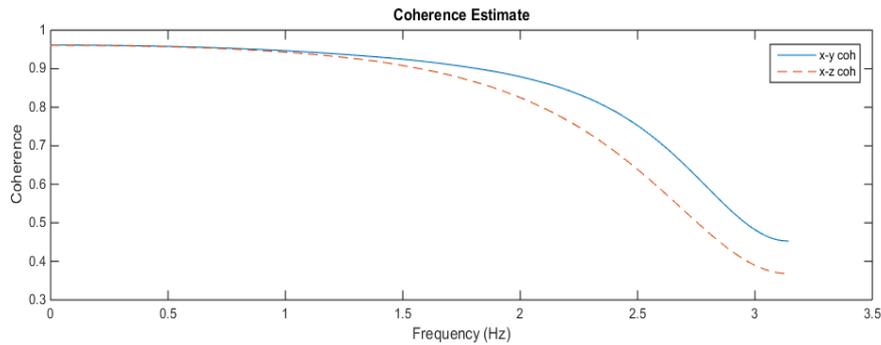

Fig. 3. Coherence estimates

The proposed approach is depicted in Fig. 4. As shown, the approach consists mainly of two phases: a learning phase and a prediction phase. The steps of the two phases can be summarized as follows:
1) A User_Profile matrix is generated, where the entry in row $i$, column $j$ is a value from 0 to 5, indicating the rating that user $i$ has provided for item $j$, with 0 indicating unrated items.
2) The User_Profile matrix is randomly divided into five disjoint partitions of nearly equal sizes. The following steps are repeated five times; each time, four of the partitions are used as training data, whereas the fifth partition is left for testing.
3) SOM is applied to the training data, and the output of the SOM is mapped back to the input space (inputs attached to each neuron are specified.)
4) K-means is applied to the mapped output of the SOM to group them into $k$ clusters.
5) In each of the five testing subsets, 10 ratings from each user are removed.
6) The closest cluster to each user in the testing subset is found.
7) FFT is applied to the columns (items) in the cluster to which the user belongs, and the spectrum of each item in that cluster is generated.
8) The coherence similarity measure is applied to each cluster to find the similarity between items based on their spectra; then, the top N similar items in each cluster are found.

9) Using the top N similar items, removed ratings are predicted.
10) Evaluation of the prediction quality is calculated for each user in the testing subset.
11) The average prediction quality is calculated for the five testing subsets using MAE, precision and recall.

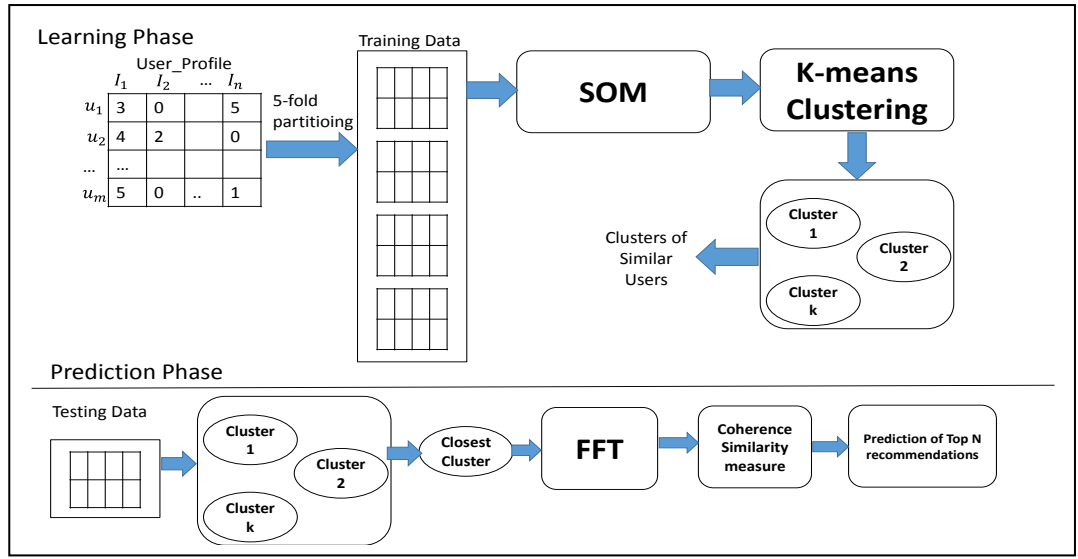

Fig. 4. The proposed approach

Note that, as shown in Fig. 4 and according to step (2) of the proposed approach, to test whether the built model has good generalization power when applied to unseen data, a portion of the available data is used for training, and the remainder is used for testing. One common method for dividing a dataset into training and test data is k-fold cross validation. In this method, k experiments are performed: k-1 folds are used for training and the remaining fold is used for testing [31]. This method guarantees that all the data points are used in both the training and testing. The average performance measures are then computed. In the performed experiments, we use k = 5; therefore, in each iteration, 20% of the data are used for testing and 80% is used as the training data.

## 4. 2 THE PREDICTION

To predict the missing ratings, we use (11), which is similar to a simple weighted average [32] except that only the top N similar items are summed:

$$P_{a,i} = \frac{\sum_{n \in N} r_{a,n} S_{i,n}}{\sum_{n \in N} S_{i,n}}, \qquad (11)$$

where $r_{a,n}$ is the rating of user $a$ on item $n$, and $S_{i,n}$ is the similarity between item $i$ and item $n$. Note that, for a cold start, where no ratings have been provided (i.e., $r_{a,n} = 0$ for all items $n$), (12) can be used instead of (11):

$$P_{a,i} = \frac{\sum_{j=1}^{|U|} r_{i,j}}{|U|}, \qquad (12)$$

where $|U|$ represents the number of users in the cluster closest to user $a$. In this case, the closest cluster will be the one with the smallest number and smallest values of provided ratings.

## 5. Experimental Study

### 5.1 Used Data

The most commonly used datasets in related research are the MovieLens datasets. This work uses the MovieLens 100K, which consists of 100,000 ratings of 1,682 movies by 943 users on a 1 to 5 scale. The data were cleaned, and users who rated fewer than 20 movies were removed. The dataset also includes simple user demographic information (age, gender, occupation, and zip code).

### 5.2 Evaluation metrics

Several metrics are used to evaluate the presented approach. Among the commonly available metrics for the performance evaluation of CF, one popular and common metric is the Mean Absolute Error (MAE) [33]. The MAE calculates the average deviation between predicted ratings and real ratings; hence, small values indicate better results. The MAE can be calculated using (13)

$$MAE = \sum_{\forall i,j} \frac{|P_{ij} - r_{ij}|}{n}, \qquad (13)$$

where $P_{ij}$ is the predicted rating for user $i$ of item $j$, $r_{ij}$ is the actual rating, and $n$ is the total number of items in the testing data.

In addition, to test the effectiveness of the proposed similarity measure in comparison to other similarity measures, recall and precision are used. Recall represents the ratio of the relevant recommended items to relevant items. On the other hand, precision is the ratio of relevant items to recommended items. These two measures should be as high as possible, although increasing one of them decreases the other [34]. A common measure that conveys the balance between both values is the F1 measure, which is the harmonic mean of precision and recall, calculated by solving (14) [35].

$$F1 = \frac{2 x Precision x Recall}{Precision + Recall}. \qquad (14)$$

Furthermore, to evaluate how the proposed approach deals with sparse data, ratings are randomly removed from the used dataset to generate different various sparsity levels. The sparsity level is denoted by $\tau$, which is calculated as follows (8):

$$\tau = \frac{R}{M*N} * 100 , \qquad (15)$$

where $R$, $M$, and $N$ are the numbers of ratings, users, and items, respectively. The sparsity levels used in the experiments are 0.18, 0.1, and 0.05.

**5.3 Results and Analysis**

**5.3.1 Performance of the clustering stage**

To evaluate the effectiveness of the proposed method, we performed several experiments. In the first experiment, we applied k-means clustering directly to the User_Profile matrix. Euclidean distance was used in the implemented k-means algorithms. The performance of the clustering algorithm was evaluated using the silhouette [36] curve for different values of k (10, 20, 30, and 40). The silhouette curve displays a measure of how close each point in one cluster is to points in neighboring clusters. This measure ranges from +1, which indicates points that are very distant from neighboring clusters, to 0, indicating points that are not distinctly in one cluster or another, to −1, indicating points that are assigned to the wrong cluster. Moreover, the vertical height of each cluster is proportional to the number of data points in each cluster. As shown in Fig. 5, many data points have negative silhouette values even when the number of clusters varies for different $k$ values. Moreover, as $k$ increases, the number of wrongly classified data points also increases. This indicates that clustering using k-means is inadequate.

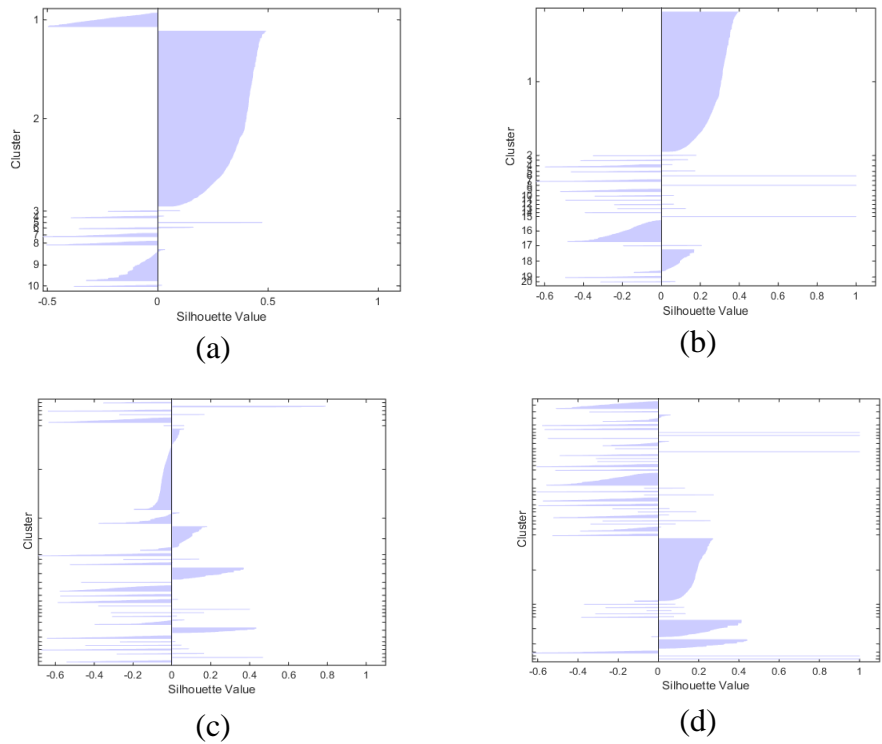

Fig. 5. Silhouette curves of k-means clustering results when different values of k. (a) k =10, (b) k = 20, (c) k = 30, and (d) k = 40

In the second performed experiment, to investigate whether SOM can enhance the clustering results and, hence, the overall results, SOM is applied first, followed by k-means, to cluster the resulting neurons after converting them back to the input space. The output of an implemented SOM is a map that forms a compressed representation of the input space. This representation reflects both the relative density of the input vectors in that space and a two-dimensional compressed representation of the input-space topology. In the experiments, the constructed SOM consists of one hidden layer with 10×10 neurons organized in a grid.

To assess the performance of the two clustering algorithms (SOM followed by k-means), plots similar to those in Fig. 5 are used. As shown in Fig. 6, the performance of the clustering algorithm with respect to the resulting topology of the used SOM is enhanced in comparison to the first approach, where only k-means was used. This is shown by the decreased number of erroneously classified data points for different values of k compared to those presented in Fig. 5. Note also that, beyond inadequate clustering, the process of clustering using k-means only was also more time consuming (the calculation time quadrupled on the same machine) in comparison with k-means followed by SOM. This result was surprising, as we expected that adding another clustering stage in addition to k-means would be more time consuming. However, due to the ability of SOM to find the hidden similarities in the input data, finding similar groups of users is apparently easier when SOM is applied first. Hence, finding similarities is more efficient using SOM followed by k-means as clustering engines.

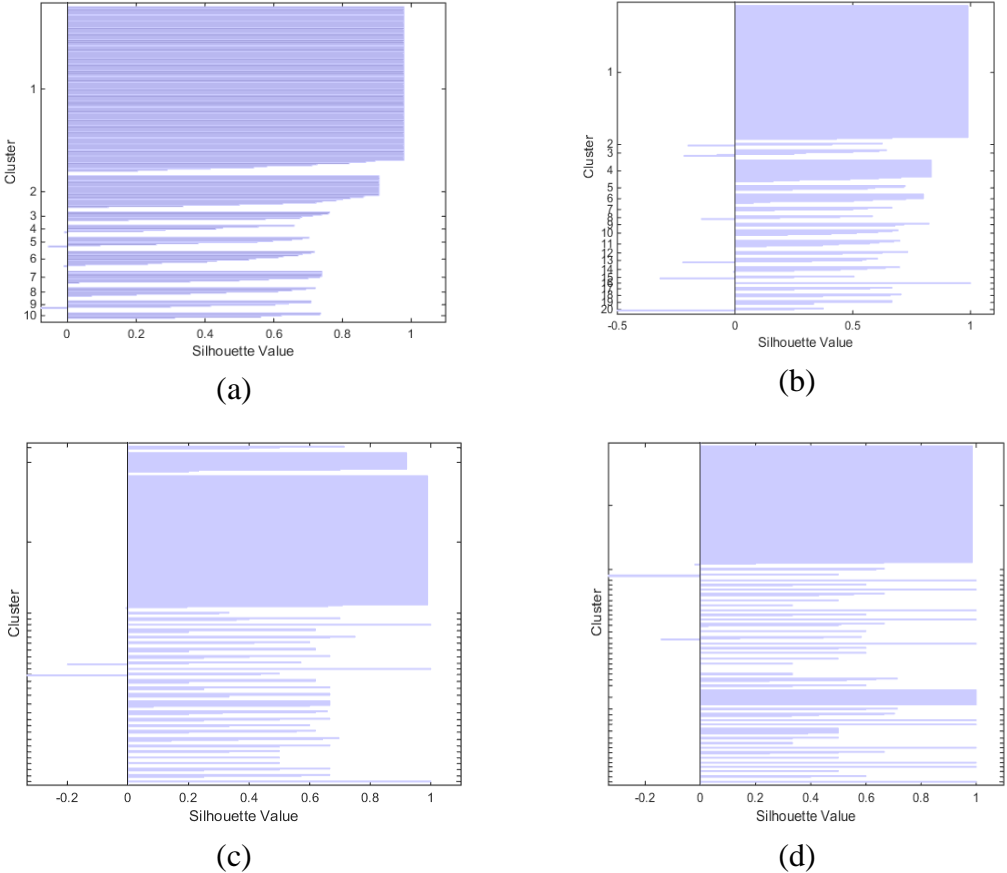

Fig. 6. Silhouette curve for SOM followed by k-means using different values of k. (a) k =10, (b) k = 20, (c) k = 30, and (d) k = 40

**5.3.2 Performance of the proposed similarity measure**

To evaluate the effectiveness of the proposed similarity measure for different sparsity values and to compare it with state-of-the-art measures, the proposed approach was implemented and tested several times while keeping all design parameters unchanged; only the similarity measure was changed. Similarity measures that were implemented and compared with the proposed method (Cohr) include PCC, Jaccard, MSD, and JMSD.

The performance of the proposed approach varies based on two main factors: the number of clusters of similar users (K) and the number of top similar items (N). The following sections, describe in detail how these two factors affect the performance of the proposed approach when using Cohr compared to similar approaches using PCC, Jaccard, MSD, and JMSD.

**5.3.2.1 The effect of varying K**

To test the effectiveness of the proposed approach as K varies, experiments were performed in which K changed from 10 to 55 while N was kept constant at 100. As shown in Table 1, increasing the value of K has a negative effect on the MAE values of Cohr compared with other measures for different sparsity values (0.18, 0.1, and 0.05). Note that although the performance of the proposed approach when Cohr is used decreases as the sparsity level increases, Cohr is more robust and shows better MAE values than the approaches that employ other measures.

Table 1. The effect of K on the MAE values for different sparsity values.

| K | Sparsity=0.18 | | | | | Sparsity=0.1 | | | | | Sparsity=0.05 | | | | |
|---|---|---|---|---|---|---|---|---|---|---|---|---|---|---|---|
| | Cohr | PCC | Jaccard | MSD | JMSD | Cohr | PCC | Jaccard | MSD | JMSD | Cohr | PCC | Jaccard | MSD | JMSD |
| 10 | **0.46** | 0.49 | 0.54 | 0.56 | 0.58 | **0.48** | 0.53 | 0.58 | 0.60 | 0.51 | **0.53** | 0.59 | 0.63 | 0.60 | 0.57 |
| 15 | **0.46** | 0.52 | 0.49 | 0.51 | 0.53 | **0.48** | 0.55 | 0.53 | 0.55 | 0.51 | **0.53** | 0.61 | 0.58 | 0.60 | 0.57 |
| 20 | **0.49** | 0.52 | 0.56 | 0.57 | 0.56 | **0.51** | 0.55 | 0.59 | 0.61 | 0.54 | **0.56** | 0.61 | 0.64 | 0.63 | 0.60 |
| 25 | 0.53 | **0.50** | 0.56 | 0.58 | 0.59 | 0.55 | **0.54** | 0.60 | 0.62 | 0.58 | **0.60** | 0.60 | 0.65 | 0.67 | 0.64 |
| 30 | 0.55 | **0.51** | 0.54 | 0.56 | 0.64 | 0.57 | **0.55** | 0.58 | 0.60 | 0.60 | 0.62 | **0.61** | 0.63 | 0.69 | 0.66 |
| 35 | 0.58 | 0.57 | **0.55** | 0.57 | 0.68 | **0.60** | 0.61 | 0.59 | 0.61 | 0.63 | 0.65 | 0.67 | 0.64 | 0.72 | 0.69 |
| 40 | 0.57 | 0.59 | **0.55** | 0.59 | 0.73 | **0.59** | 0.63 | 0.58 | 0.63 | 0.62 | 0.64 | 0.69 | **0.63** | 0.71 | 0.68 |
| 45 | **0.59** | 0.62 | 0.62 | 0.66 | 0.74 | **0.61** | 0.65 | 0.66 | 0.70 | 0.64 | **0.66** | 0.71 | 0.71 | 0.73 | 0.70 |
| 50 | **0.58** | 0.61 | 0.64 | 0.68 | 0.78 | **0.60** | 0.65 | 0.68 | 0.72 | 0.63 | **0.65** | 0.71 | 0.73 | 0.72 | 0.69 |
| 55 | **0.58** | 0.62 | 0.60 | 0.70 | 0.79 | **0.60** | 0.65 | 0.64 | 0.74 | 0.63 | **0.65** | 0.71 | 0.69 | 0.72 | 0.69 |

As another evaluation of the effectiveness of proposed measure, we compared the F1 measure of the proposed similarity measure to similar measures using different sparsity values. Table 2 shows how the F1 values change for different sparsity values. Similarly, the table shows that the proposed similarity measure achieves higher F1 values compared with other similarity measures.

Table 2. The effect of K on the F1 values for different sparsity values.

| K | Sparsity=0.18 | | | | | Sparsity=0.1 | | | | | Sparsity=0.05 | | | | |
|---|---|---|---|---|---|---|---|---|---|---|---|---|---|---|---|
| | Cohr | PCC | Jaccard | MSD | JMSD | Cohr | PCC | Jaccard | MSD | JMSD | Cohr | PCC | Jaccard | MSD | JMSD |
| 10 | **0.61** | 0.43 | 0.38 | 0.49 | 0.50 | **0.49** | 0.48 | 0.40 | 0.39 | 0.49 | **0.44** | 0.32 | 0.35 | 0.34 | 0.39 |
| 15 | **0.64** | 0.47 | 0.41 | 0.53 | 0.53 | **0.53** | 0.52 | 0.43 | 0.43 | 0.52 | **0.47** | 0.34 | 0.38 | 0.38 | 0.42 |
| 20 | **0.66** | 0.50 | 0.43 | 0.55 | 0.55 | **0.55** | 0.54 | 0.45 | 0.45 | 0.54 | **0.49** | 0.35 | 0.40 | 0.40 | 0.44 |
| 25 | **0.68** | 0.52 | 0.45 | 0.57 | 0.57 | **0.57** | 0.56 | 0.46 | 0.47 | 0.56 | **0.50** | 0.36 | 0.41 | 0.42 | 0.46 |
| 30 | **0.69** | 0.54 | 0.46 | 0.58 | 0.58 | **0.58** | 0.58 | 0.48 | 0.48 | 0.57 | **0.51** | 0.37 | 0.43 | 0.43 | 0.47 |
| 35 | **0.70** | 0.56 | 0.47 | 0.60 | 0.59 | **0.60** | 0.59 | 0.49 | 0.50 | 0.58 | **0.52** | 0.38 | 0.44 | 0.45 | 0.48 |
| 40 | **0.71** | 0.57 | 0.48 | 0.61 | 0.60 | **0.61** | 0.61 | 0.50 | 0.51 | 0.59 | **0.53** | 0.38 | 0.45 | 0.46 | 0.49 |
| 45 | **0.72** | 0.58 | 0.49 | 0.62 | 0.61 | **0.62** | 0.62 | 0.50 | 0.52 | 0.60 | **0.54** | 0.39 | 0.45 | 0.47 | 0.50 |
| 50 | **0.73** | 0.59 | 0.50 | 0.63 | 0.62 | **0.63** | 0.63 | 0.51 | 0.53 | 0.61 | **0.54** | 0.40 | 0.46 | 0.48 | 0.51 |
| 55 | **0.74** | 0.60 | 0.51 | 0.63 | 0.63 | **0.63** | 0.63 | 0.52 | 0.53 | 0.62 | **0.55** | 0.40 | 0.47 | 0.48 | 0.52 |

### 5.3.2.2 The effect of varying N

A similar analysis was performed to study the effect of N on the obtained results, where K was kept constant at 55. Table 3 shows how the values of MAE change as N is varied from 10 to 100 for different sparsity values (0.18, 0.1, and 0.05). As shown in the table, the Cohr outperforms similar measures for different sparsity levels. Moreover, Cohr seems to be more robust to the sparsity problem because its MAE values are only slightly affected by the changes in the sparsity values compared to the other measures. This occurs because the proposed measure can find the similarity between sequences of data points in a global manner, even if most of the values of the compared data points are equal to zero.

Table 3. The effect of N on the MAE values for different sparsity values.

| N | Sparsity=0.18 | | | | | Sparsity=0.1 | | | | | Sparsity=0.05 | | | | |
|---|---|---|---|---|---|---|---|---|---|---|---|---|---|---|---|
| | Cohr | PCC | Jacc | MSD | JMSD | Cohr | PCC | Jacc | MSD | JMSD | Cohr | PCC | Jacc | MSD | JMSD |
| 10 | **0.45** | 0.63 | 0.58 | 0.49 | 0.61 | **0.47** | 0.65 | 0.68 | 0.70 | 0.61 | **0.67** | 0.85 | 0.78 | 0.70 | 0.81 |
| 20 | **0.45** | 0.64 | 0.59 | 0.52 | 0.61 | **0.50** | 0.68 | 0.69 | 0.73 | 0.61 | **0.70** | 0.88 | 0.79 | 0.73 | 0.81 |
| 30 | **0.46** | 0.64 | 0.59 | 0.52 | 0.61 | **0.51** | 0.70 | 0.69 | 0.73 | 0.61 | **0.71** | 0.90 | 0.69 | 0.73 | 0.81 |
| 40 | **0.47** | 0.62 | 0.57 | 0.50 | 0.59 | **0.50** | 0.62 | 0.67 | 0.71 | 0.59 | **0.70** | 0.82 | 0.77 | 0.71 | 0.89 |
| 50 | **0.47** | 0.64 | 0.59 | 0.51 | 0.63 | **0.52** | 0.64 | 0.69 | 0.72 | 0.63 | **0.62** | 0.84 | 0.79 | 0.72 | 0.83 |
| 60 | **0.46** | 0.62 | 0.55 | 0.49 | 0.59 | **0.56** | 0.72 | 0.66 | 0.70 | 0.59 | **0.66** | 0.82 | 0.76 | 0.70 | 0.89 |
| 70 | 0.51 | 0.63 | 0.58 | **0.50** | 0.60 | **0.57** | 0.73 | 0.66 | 0.71 | 0.60 | **0.67** | 0.83 | 0.76 | 0.71 | 0.90 |
| 80 | **0.52** | 0.62 | 0.57 | **0.52** | 0.59 | **0.59** | 0.72 | 0.67 | 0.73 | 0.59 | **0.69** | 0.92 | 0.77 | 0.73 | 0.89 |
| 90 | 0.52 | 0.64 | 0.59 | **0.51** | 0.61 | **0.60** | 0.74 | 0.69 | 0.72 | 0.61 | **0.70** | 0.94 | 0.79 | 0.72 | 0.81 |
| 100 | **0.50** | 0.63 | 0.58 | 0.52 | 0.60 | **0.60** | 0.73 | 0.68 | 0.73 | 0.60 | **0.70** | 0.93 | 0.78 | 0.73 | 0.80 |

## 6. CONCLUSION AND FUTURE WORK

In this paper, a new similarity measure is proposed as an approach for CF. The proposed approach has several advantages. First, the combination of two powerful clustering approaches, namely, SOM and k-means, yields high performance because SOM has a powerful ability to find hidden similarities in the input data, while k-means is a simple clustering algorithm that finds similar data efficiently and groups them into clusters. Moreover, the combination of these two algorithms enables us to build a scalable approach because it finds the users most similar to the active user without accessing the entire User_Profile matrix without sacrificing prediction quality. Second, a new similarity measure based on the spectral domain is defined. The rationale behind the proposed measure is the ability of spectral features to be easily calculated and analyzed using DSP techniques. The main tools used in the presented approach include the FFT and spectral coherence. Coherence takes the global similarity between users into account by considering the complete set of ratings provided by each user, which yields high performance measures. Moreover, the new measure can deal with the sparsity problem gracefully because no matrix inversion is involved. Furthermore, the calculation of the proposed measure is more time efficient (O(nlogn)) in comparison with the calculation of other similarity measures (O($n^2$)). Finally, the obtained experimental results show that the proposed measure achieves high performance scores compared with other state-of-the-art measures while alleviating the tradeoffs between prediction accuracy and scalability.

This work enables other DSP techniques to be adopted in the CF context, which is what we intend to investigate in future work. Moreover, the proposed approach will be applied to several datasets to evaluate its generalization ability.